\documentstyle[aps,prl,multicol,floats,epsf]{revtex}

\begin{document}

\draft

\wideabs{
\title{
Temperature- and magnetic-field-dependent thermal conductivity of \\
pure and Zn-doped Bi$_{2}$Sr$_{2}$CaCu$_2$O$_{8+\delta}$ single crystals}

\author{Yoichi Ando,$^1$ J. Takeya,$^1$ Yasushi Abe,$^{1,2}$ 
K. Nakamura,$^{1,3}$ and A. Kapitulnik$^4$}

\address{$^1$Central Research Institute of Electric Power
Industry, Komae, Tokyo 201-8511, Japan}
\address{$^2$Department of Physics, Science University of Tokyo, 
Shinjuku-ku, Tokyo 162-8601, Japan}
\address{$^3$Department of Energy Science, Tokyo Institute of Technology,
Nagatsuta, Yokohama 226-8502, Japan}
\address{$^4$E.L. Ginzton Laboratory, Stanford University, 
Stanford, CA 94305, USA}

\date{BSCCO-s3.tex}
\maketitle

\begin{abstract}
The thermal conductivity $\kappa$ of 
Bi$_{2}$Sr$_{2}$CaCu$_2$O$_{8+\delta}$ 
is measured in pure and Zn-doped crystals as a function of temperature
and magnetic field.
The in-plane resistivity is also measured on the identical samples.
Using these data, we make a crude estimate of the impurity-scattering
rate $\Gamma$ of the pure and the Zn-doped crystals. 
Our measurement show that the ``plateau" in the $\kappa(H)$ profile 
is not observed in the majority of our Bi-2212 crystals, 
including one of the cleanest crystals available to date.
The estimated values of $\Gamma$ for the pure and Zn-doped samples
allow us to compare the $\kappa(H)$ data with the existing theories
of the quasiparticle heat transport in $d$-wave superconductors 
under magnetic field.
Our analysis indicates that a proper inclusion of the 
quasiparticle-vortex scattering, which is expected to play the key 
role in the peculiar behavior of the $\kappa(H)$, is important for 
a quantitative understanding of 
the QP heat transport in the presence of the vortices.
\end{abstract}

\pacs{PACS numbers: 74.25.Fy, 74.62.Dh, 74.72.Hs}
}
\narrowtext

Thermal conductivity $\kappa$ of a superconductor 
is one of the few probes which 
allow us to investigate the quasiparticle (QP) density and its 
scattering rate in the superconducting (SC) state.
It is now believed that the SC state of the high-$T_c$ cuprates 
is primarily $d_{x^2-y^2}$ \cite{d-wave},
where it has been found 
that the magnetic field induces extended QPs 
whose population increases as $\sqrt{H}$ \cite{Moler,Volovik}.
Also, the QP scattering rate in the cuprates has been found to 
drop very rapidly below $T_c$ \cite{Bonn},
which causes a pronounced peak in the temperature dependence of 
$\kappa$ \cite{Krishana}.
In 1997, Krishana {\it et al.} reported an intriguing result from 
their measurement on the 
magnetic-field dependence of $\kappa$ in 
Bi$_{2}$Sr$_{2}$CaCu$_2$O$_{8+\delta}$ (Bi-2212) at temperatures 
below 20 K \cite{Ong}.
They observed, in the profile of $\kappa (H)$, a sharp break at a 
\lq\lq transition field" $H_k$ and a subsequent plateau region where
$\kappa$ does not change with magnetic field.
Krishana {\it et al.} proposed an interpretation 
that $H_k$ marks a phase 
transition from the $d_{x^2-y^2}$ state to a fully-gapped
$d_{x^2-y^2}+id_{xy}$ (or $d_{x^2-y^2}+is$) state and thus 
in the plateau region there are little thermally-excited
QPs which contribute to the heat transport.
This interpretation appears to be fundamentally related to the
high-$T_c$ mechanism and therefore attracted much attention 
both from theorists 
\cite{Laughlin,Ogata,Rama,Ghosh,Liu,Anderson,Franz,Vekhter} 
and from experimentalists 
\cite{Aubin1,Aubin2,Zeini,Taldenkov}.

An independent test of this unusual behavior of $\kappa$ 
has been reported by Aubin {\it et al.} \cite{Aubin1}; 
although the plateau-like feature was essentially reproduced in
their measurement, when the data were taken 
with field sweep up and down, Aubin {\it et al.} observed
a rather large jump in $\kappa$ upon field reversal and consequently
the $\kappa (H)$ profile had a pronounced hysteresis.
The fact that the $\kappa$ value in the ``plateau" 
depends on the history
of the applied magnetic field casts a serious doubt on the Krishana 
{\it et al.}'s interpretation.
Moreover, Aubin {\it et al.} reported 
that an {\it increase} in $\kappa$ 
with magnetic field was observed  at subkelvin temperatures, 
which strongly suggests the presence of 
a finite density of QPs at low temperatures 
and thus is incompatible with the fully-gapped state \cite{Aubin2}.
Although these newer results suggest that a novel phase transition 
in the gap symmetry is not likely to be taking place, 
the plateau in the $\kappa (H)$ profile
and the sensitiveness of $\kappa$ on the magnetic-field
history are still to be understood.
One interesting information one can draw from these experiments 
is that phonons are {\it not} scattered by vortices 
in cuprate superconductors \cite{Ong,Aubin1,Ong_new}.

Motivated by these experiments, there appeared several theories 
that try to capture the essential physics of the QP 
heat transport in the $d$-wave superconducting state.
It has become rather clear that a proper inclusion of the 
QP-vortex scattering rate is necessary for explaining the 
observed magnetic-field dependence; however, there has been no
consensus yet as to {\it how} the QP-vortex scattering is 
taken into account.  To improve our understanding of the 
QP heat transport under magnetic field, quantitative 
examinations of various theories in the light of actual data 
are indispensable.  For Bi-2212, however, previously published data
from various groups do not provide enough information;
for example, the impurity scattering rate, which is the most 
important parameter that controls the thermal conductivity behavior,
are not known for clean Bi-2212 crystals.
(Rather surprisingly, no electrical resistivity data have been 
supplied for samples which were used in recent studies of $\kappa(H)$ 
profile of Bi-2212 \cite{Ong,Aubin1,Aubin2,Taldenkov}.)

In this paper, we present results of our measurements
of $\kappa(T)$ and $\kappa(H)$ of well-characterized 
Bi-2212 crystals, together with their in-plane resistivity 
($\rho_{ab}$) data.
To look for the effect of changing impurity-scattering rate,
we measured both pure and 0.6\%-Zn-doped crystals.
The crystals used here are single domained (without any mosaic structure
nor grain boundaries) and have very good morphology, 
which were confirmed by 
x-ray analyses and the polarized-light optical microscope analysis.
In the data presented here, neither the pure sample nor the Zn-doped sample 
show any plateaus in the $\kappa(H)$ profile in the temperature and the
magnetic-field regime where the plateaus have been reported.
In fact, we have found that the plateau in the $\kappa(H)$ profile is 
not a very reproducible feature (we have thus far observed the 
plateau-like feature 
in only 2 samples out of more than 30 samples measured) 
and we have not yet conclusively sorted out what determines the 
occurrence of the plateau \cite{mosaic}; we therefore decided to show only 
the data that are representative of the majority of the samples.
Using all the data of $\rho_{ab}(T)$, $\kappa(T)$, and $\kappa(H)$, 
we try to estimate the electronic thermal conductivity $\kappa_e(T)$ 
and make a rough estimate of the impurity scattering rate $\Gamma$ 
for the pure and Zn-doped samples.
Our data offer a starting point 
for the quantitative understanding of the
QP heat transport in the $d$-wave superconductors in the magnetic fields,
where an interplay between the QP-vortex scattering and the 
QP-impurity scattering apparently plays an important role.

The single crystals of Bi-2212 are grown with a floating-zone method and
are carefully annealed and quenched to obtain a uniform oxygen content 
\cite{Ando}.  Both the pure and the Zn-doped crystals 
[Bi$_{2}$Sr$_{2}$Ca(Cu$_{1-z}$Zn$_z$)$_2$O$_{8+\delta}$] 
are tuned to the optimum doping 
by annealing at 750 $^\circ$C for 48 hours, after which the transition 
width of about 1.5 K (measured by the dc magnetic 
susceptibility measurements) 
is achieved; this indicates a high homogeneity of these crystals. 
The Zn-doped crystal contains (0.6$\pm$0.1)\% of Zn 
(namely $z$=0.006$\pm$0.001), 
which was determined by the inductively-coupled plasma spectrometry.
The zero-resistance $T_c$'s are 92.4 K 
for the pure crystals and 84.5 K for the Zn-doped crystal.
We first measure the temperature dependence of $\rho_{ab}$ using 
a standard four-probe technique, and then measure the thermal conductivity
$\kappa$ of the same sample.
The temperature dependence of $\kappa$ from 1.6 to 160 K
are measured in zero field using calibrated AuFe-Chromel thermocouples.
The precise magnetic-field dependence of $\kappa$ 
is measured with  
a steady-state technique using a small home-made thin-film heater 
and microchip Cernox thermometers.
The bottom end of the sample is anchored 
to a copper block whose temperature is
carefully controlled within 0.01\% stability and accuracy. 
Since we need to measure the change in $\kappa$ with a very high
accuracy, the small magnetic-field dependence of the thermometers
were carefully calibrated beforehand using a SrTiO$_3$ capacitance sensor 
and a high-resolution capacitance bridge.  
As a check for the measurement system and the calibrations, we first
measured the thermal conductivity of nylon and confirmed that the
reading is indeed magnetic-field independent within 0.1\% accuracy.

\begin{figure}[b!]
\epsfxsize=1.0\columnwidth
\centerline{\epsffile{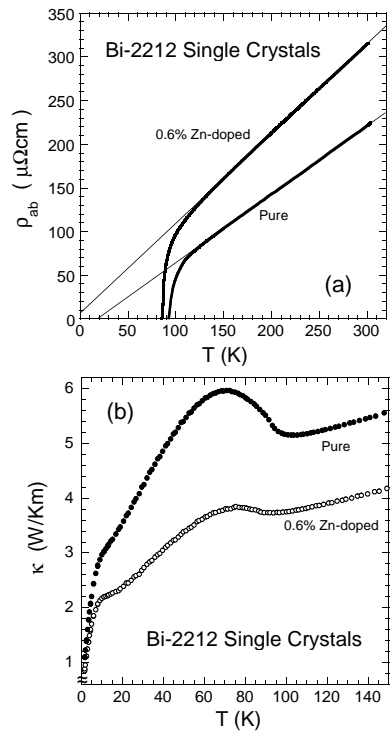}}
\vspace{-1cm}
\caption{$T$ dependence of (a) $\rho_{ab}$ and (b) $\kappa$ 
of pure and 0.6\%-Zn-doped Bi-2212 crystals in zero field.  
The thin solid lines are $T$-linear fits to the 
$\rho_{ab}(T)$ data.}
\label{fig1}
\end{figure}

Figure 1(a) shows the $\rho_{ab}(T)$ data of the pure and Zn-doped samples.
If we define the residual resistivity $\rho_0$ by the 
extrapolation of the $T$-linear resistivity to 0 K, we obtain 
slightly negative $\rho_0$ for the pure sample; this is always the 
case for the cleanest Bi-2212 crystals grown in our laboratory.
As is expected, the Zn-doped sample gives larger $\rho_0$, which is
about 10 ${\rm \mu \Omega cm}$.
We note that the uncertainty in the absolute magnitude of $\rho_{ab}$ and 
$\kappa$ in our measurements are less than $\pm$5\% in this work. 
This is achieved by determining the sample thickness, which is usually
the main source of the uncertainty, by measuring the weight of the sample
with 0.1 $\mu$g resolution.  We used relatively long samples here
(the length of the pure and the Zn-doped samples were 4.5 mm 
and 6.5 mm, respectively), with which the errors in estimating 
the voltage contact separation and the thermocouples separation 
are less than $\pm$5\%.
It should also be noted that a sophisticated technique to make the 
current contacts uniformly on the side faces of the crystals are 
crucial for reliably measuring $\rho_{ab}$ of Bi-based cuprates.

Figure 1(b) shows the temperature dependence of $\kappa$ 
of the two samples in zero field.  
To our knowledge, the size of the peak in $\kappa(T)$ of this 
pure sample is the largest ever reported for Bi-2212 
(the enhancement from the minimum near $T_c$ to the peak is 16\%).
This implies that the pure crystal reported here is among 
the cleanest Bi-2212 crystals available to date.
The Zn-doped sample shows not only a smaller magnitude of $\kappa$
but also a significantly suppressed peak in $\kappa(T)$
below $T_c$; this is caused by a larger impurity scattering rate
(which limits the enhancement of the QP mean free length below $T_c$)
and is consistent with the Zn-doping effect in 
YBa$_2$Cu$_3$O$_{7-\delta}$ (YBCO) \cite{Hirschfeld,Chiao}.

\begin{figure}[b!]
\epsfxsize=1.0\columnwidth
\centerline{\epsffile{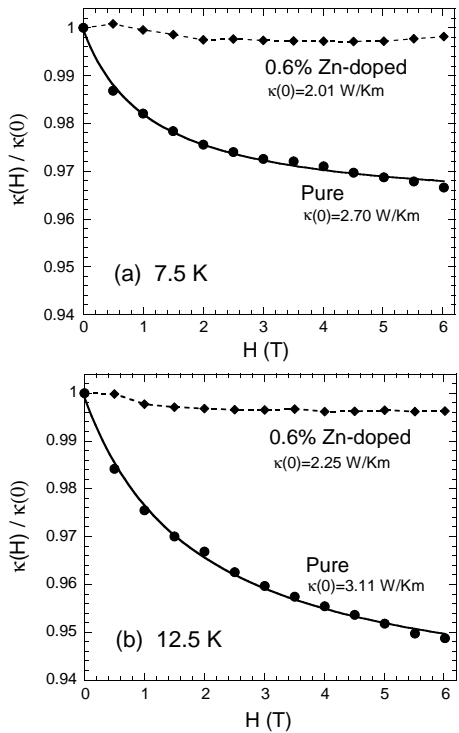}}
\vspace{-0.3cm}
\caption{Normalized $\kappa (H)$ profiles of the pure and Zn-doped
samples at (a) 7.5 K and (b) 12.5 K. 
The solid lines are the fits to Eq. (1) with:
(a) $\kappa_{ph}$=2.60 W/Km, $\kappa_e$=0.10 W/Km, and $p$=0.88 for the
7.5 K data, and (b) $\kappa_{ph}$=2.91 W/Km, $\kappa_e$=0.20 W/Km, and 
$p$=0.53 for the 12.5 K data.}
\label{fig2}
\end{figure}

Figure 2 shows $\kappa (H)$ profiles of the two samples 
at 7.5 and 12.5 K.  The data are normalized by the zero-field 
value of $\kappa$.
Neither of these samples show any plateau-like feature below 6 T.
(The $\kappa(H)$ data of the Zn-doped sample is almost flat within
our sensitivity, but we do not call it a plateau.) 
Note that Ref. \cite{Ong} reported the plateaus to be observed above 
0.9 T at 7.5 K and above 2.6 T at 12.5 K, which are well within the
range of our experiment.
The data in Fig. 2 were taken in the field-cooled (FC) 
procedure, in which the sample is cooled in the presence of 
a magnetic field, and therefore the magnetic induction in the sample 
remains homogeneous. 
The FC data are expected to give information which is free from
complications due to vortex pinning, while the zero-field-cooled (ZFC)
data (for which the sample is cooled in zero field and the 
magnetic field is swept at a constant temperature) 
are subject to such complications.
We emphasize that the FC data should be examined to 
look for the intrinsic effect of vortices on the QP transport. 
All the previously published data on the plateau 
\cite{Ong,Aubin1,Aubin2,Taldenkov} are, however, ZFC data \cite{note}.

It has been proposed 
that the magnetic-field dependence
of $\kappa$ of the high-$T_c$ cuprates are described by
\begin{equation}
\kappa(H,T)= \frac{\kappa_e(T)}{1+p(T)|H|} + \kappa_{ph}(T),
\label{eq}
\end{equation}
where $\kappa_e$ is the electronic part of $\kappa$ in zero field 
and $\kappa_{ph}$ is the phonon part \cite{note2}.
Equation (1) was proposed first by Yu {\it et al.} \cite{Yu} 
and later by Ong and co-workers \cite{Krishana,Ong_new,Krishana_new}. 
This expression utilizes the finding that the phonon thermal
conductivity of the cuprates is independent of the magnetic field.
The parameter $p(T)$ is proportional to the zero-field value 
of the QP mean free path.
We found that the $\kappa(H)$ data of our clean samples are 
reasonably well described by Eq. (1).
The solid lines in Fig. 2 (a) and (b) are fits of the data to 
Eq. (1).  The fitting parameters suggest that the 
phononic contribution to $\kappa$ is as large as 96\% and 93\%
at 7.5 and 12.5 K, respectively; this is essentially a reflection
of the fact that the changes in $\kappa$ with the magnetic field
are very small at these temperatures.
Ong {\it et al.} reported \cite{Ong_new} 
that $p(T)$ is about 2.3 T$^{-1}$ 
for underdoped YBCO at 7.5 K, while it is 0.88 T$^{-1}$ for our pure
Bi-2212 at 7.5 K.  
This is an indication that the QP transport
is dirtier in Bi-2212 compared to YBCO; we will elaborate on 
these $p(T)$ values later.
The magnetic field dependence of $\kappa$ of the Zn-doped sample
is too small to make a reliable fit to Eq. (1), but one can infer
that the ratio of the phononic contribution in the Zn-doped sample 
is even larger than in the pure sample.

In the normal state, the electronic thermal conductivity $\kappa_e$
and the electrical conductivity $\sigma$ are related by 
$\kappa_e/T$ = $L\sigma$, where $L$ is called the Lorenz number.
In simple metals, $L$ is usually constant at high temperatures 
(Wiedemann-Franz law) and the free-electron model gives 
$L_0$ = 2.44$\times$10$^{-8}$ W$\Omega$/K$^2$.
When the electron-electron correlation becomes strong, 
$L$ becomes smaller than the free electron value $L_0$; 
for YBCO, $L$ has been 
estimated to be around 1.0$\times$10$^{-8}$ W$\Omega$/K$^2$ 
near $T_c$ \cite{Hirschfeld,Zhang}.
Using this value of $L$ and the $\rho_{ab}$ data 
of our crystals, we can roughly estimate $\kappa_e$ and $\kappa_{ph}$ 
above $T_c$. 
Such estimation for the pure sample 
gives $\kappa_e$ and $\kappa_{ph}$ values of about 1.6 and 3.6 W/Km,
respectively, at 120 K. 
(120K is the lower bound for the temperature range where 
the effect of the superconducting fluctuations on $\rho_{ab}$ 
is negligible.)
It has already been established that $\kappa_{ph}$ does not change 
much just below $T_c$ and that it is mostly $\kappa_e$ that causes 
the peak \cite{Krishana,Yu2}.  
We can therefore infer (from the $\kappa_e$ value
at 120 K and the total $\kappa$ at the peak) that $\kappa_e$ 
at the peak is about 1.7 times larger than in the normal state.
In the same manner, we can estimate for the Zn-doped sample
$\kappa_e$ and $\kappa_{ph}$ to be about 1.0 and 2.9 W/Km at 120 K,
respectively, and the enhancement of $\kappa_e$ at the
peak is inferred to be a factor of 1.2.
This analysis indicates that the effect of 0.6\%-Zn doping is weaker 
for $\kappa_{ph}$ (which is decreased by 19\% at 120 K upon the 
0.6\%-Zn doping) than for $\kappa_e$ (which is decreased by 38\% at 
120 K).

Comparison of the inferred behaviors of $\kappa_e$ below $T_c$ 
with the theoretical calculations of Ref. \cite{Hirschfeld} 
gives us the idea of the magnitude of the impurity-scattering rate 
$\Gamma$ in our samples.  
As is discussed above, $\kappa_e$ can be inferred to be 
about 1.7 times enhanced at the peak; 
comparison of this enhancement factor with the calculations 
for various $\Gamma/T_c$ values \cite{Hirschfeld} 
suggests that $\Gamma/T_{c}$ of our pure sample is about 0.05. 
The position of the peak (at $T/T_c \sim$ 0.7) 
is also consistent with the theoretical calculation for 
$\Gamma/T_c \sim$ 0.05.
Similarly, the inferred enhancement factor of $\kappa_e(T)$ of 
the Zn-doped sample suggests $\Gamma/T_{c}$ to be about 0.2.
Although these estimates are very crude,
the estimated values of $\Gamma/T_{c}$ of our Bi-2212 samples 
imply that the scattering rate increases by 
$\Gamma/T_{c} \sim$ 0.25 per 1\% of Zn, which is of the same 
order of magnitude as that for YBCO \cite{Hirschfeld,Taillefer}.
The estimated $\Gamma/T_{c}$ of our pure Bi-2212 is 
still notably larger than that for pure YBCO, for which 
$\Gamma/T_{c}$ has been estimated to be $\sim$0.01 
\cite{Hirschfeld,Taillefer}.
This is essentially a reflection of the fact that the peak in 
$\kappa(T)$ of Bi-2212 is much smaller compared to that of YBCO,
and is probably caused by the intrinsic disorder of the crystalline
lattice of Bi-2212 (the modulation structure along the $b$ axis).

The above mentioned difference in $\Gamma/T_{c}$ between our 
pure Bi-2212 ($\Gamma/T_c \sim$ 0.05) and pure YBCO
($\Gamma/T_c \sim$ 0.01) indicates that the QP mean free path 
in zero field, $l_0$, is roughly 5 times longer in pure 
YBCO compared to that in pure Bi-2212. 
This observation yields a useful information on the QP scattering
cross section of the vortices, $\sigma_{tr}$, in Bi-2212.
It was discussed in Refs. \cite{Ong_new} and \cite{Krishana_new} 
that parameter $p(T)$ appearing in Eq. (1) can be expressed as 
$p(T) = l_0 \sigma_{tr}/\phi_0$.
(Note that $\sigma_{tr}$ is the scattering cross section in 
two dimensions.) 
As is already discussed, the $p(T)$ value at 7.5 K is 0.88 T$^{-1}$
for our pure Bi-2212, while it is 2.3 T $^{-1}$ for YBCO; 
given the indication that $l_0$ is roughly 5 times longer
in YBCO, the ratio of the $p(T)$ values of the two systems suggests 
that $\sigma_{tr}$ should be approximately a factor of 2 larger 
in Bi-2212.  Since $\sigma_{tr}$ has been estimated to be 9 nm 
for YBCO \cite{Ong_new}, we obtain $\sigma_{tr} \sim$ 18 nm 
for Bi-2212.  The origin of this difference in $\sigma_{tr}$
between the two systems might be related to the difference in the 
structure of the vortex lines and should be a subject of future 
studies.  In any case, the estimate of $\sigma_{tr} \sim$ 18 nm 
gives $l_0 \sim$ 0.1 $\mu$m for pure Bi-2212 at 7.5 K.

Now let us discuss the observed magnetic-field dependence of $\kappa$
in conjunction with the estimated values of $\Gamma/T_{c}$ for
each sample. 
K\"ubert and Hirschfeld (KH) calculated the magnetic-field dependence 
of $\kappa_e$ that comes from the QP's Doppler shift around the 
vortex \cite{Kubert}.
Good agreements between the KH theory and experiments have been 
reported for very low temperatures, where an increase of 
$\kappa$ with magnetic field has been observed 
\cite{Aubin2,Chiao,Chiao2}.
The numerical calculation by KH show that $\kappa(H)$ is already 
an increasing function of $H$ at $T$=0.2$T_c$ for a dirty case, 
$\Gamma/T_c$=0.1; this is clearly in disagreement with our
data and indicates the necessity of an inclusion of the 
QP-vortex scattering into the calculation.  

The theory proposed by Franz \cite{Franz} tries to 
incorporate the effect of QP-vortex scattering;
heuristically, Franz supposed that the QP-impurity scattering and 
the QP-vortex scattering are separable and additive.
In this theory, 
the zero-field scattering rate $\sigma_0$ (which is a sum of the
impurity scattering rate and the inelastic scattering rate)
is directly related to the total change in $\kappa$ with the
magnetic field, $\Delta\kappa$, which is expressed as 
$\Delta\kappa$ = $(2.58T/\sigma_0 - 1)\kappa_{00}$ for $\sigma_0 <T$,
and as 
$\Delta\kappa$ = $(2.15T/\sigma_0)^2\kappa_{00}$ for $\sigma_0 >T$.
In these expressions, $\kappa_{00}$ is the universal 
thermal conductivity \cite{Taillefer,Lee}.
Assuming that the inelastic scattering is negligible 
($\sigma_0 \approx \Gamma$) at 7.5 K, we can estimate 
$\Delta\kappa$ for the pure and Zn-doped samples using these
equations.  The results are $\Delta\kappa \approx$ 0.4 W/Km for 
the pure sample and $\Delta\kappa \approx$ 0.1 W/Km for 
the 0.6\%-Zn-doped sample.
(In the calculation, we used $\kappa_{00}/T$ = 0.015 W/K$^2$m 
which is reported for Bi-2212 \cite{Chiao2}.)
When we turn to our data, the values of $\Delta\kappa$ actually 
measured in our experiments are much smaller; at 7.5 K, 
$\Delta\kappa$ up to 6 T is 0.086 and 0.008 W/Km 
for the pure and the Zn-doped samples, respectively.
Clearly, the theory overestimates $\Delta\kappa$ and the 
overestimation is particularly large for the Zn-doped sample.
This comparison implies that the rather simple treatment of 
the separable QP-impurity scattering and 
the QP-vortex scattering is not accurate enough, 
particularly when the impurity scattering rate becomes 
comparable to the vortex scattering rate.

Vekhter and Houghton (VH) \cite{Vekhter} have recently proposed a 
theory which explicitly considers the interplay between the 
vortex-lattice scattering and the impurity scattering.
Unfortunately, numerical calculations for the magnetic-field 
dependence of $\kappa$ at intermediate temperatures are 
available only for a very clean case ($\Gamma/T_c$=0.006)
\cite{Vekhter},
and we cannot make a direct comparison to our data.
Qualitatively, however, the theory predicts that a steep drop in 
$\kappa(H)$ at low fields is expected only when $T$ is larger
than $\Gamma$.  The estimated $\Gamma/T_c$ values of our samples 
suggest that the condition $T > \Gamma$ is satisfied in the 
pure sample, but is not satisfied in the 0.6\%-Zn-doped sample
in the temperature range of Fig. 2.

In summary, the temperature and magnetic-field dependences of the
thermal conductivity $\kappa$ are measured in 
pure and Zn-doped Bi-2212 crystals that are well-characterized.  
The $\rho_{ab}(T)$ data 
taken on the identical samples are used for extracting the 
electronic thermal conductivity $\kappa_e$ above $T_c$.
The temperature and magnetic-field dependences of 
$\kappa$ clearly reflect the 
difference in the impurity-scattering rate $\Gamma$ in the crystals.
It is found that in the majority of our Bi-2212 crystals 
(including one of the cleanest crystals available to date)
the ``plateau" in the $\kappa(H)$ profile is not observed 
and the $\kappa(H)$ profile around 10 K are reasonably well described 
by Eq. (1), which was originally proposed for YBCO. 
We estimate $\Gamma$ of our crystals by comparing the behaviors 
of $\kappa_e$ below $T_c$ with the calculation by Hirschfeld
and Putikka \cite{Hirschfeld}.
The estimated values of $\Gamma$ for the pure and Zn-doped samples
allow us to compare the $\kappa(H)$ data with the existing theories
of the QP heat transport in $d$-wave superconductors under magnetic
field.

We thank H. Aubin, K. Behnia, M. Franz, T. Hanaguri, A. Maeda, 
Y. Matsuda, and N. P. Ong for fruitful discussions.

%


\begin{references}
\bibitem{mosaic}
We have found, however, that at least one of the samples 
which showed a plateau-like feature in $\kappa(H)$ had a 
mosaic structure.
\bibitem{note}
We performed the ZFC measurements on some of the clean samples that do 
not show plateaus in the FC measurements; our conclusion is that the 
absence of the plateau in our data is not due to the difference in the 
measurement procedure.
\bibitem{note2}
Equation (1) is expected to be valid when the electronic specific heat
$c_e$ is independent of $H$ and the QP-vortex scattering rate is proportional
to $H$. 
The former condition is roughly satisfied in our temperature range,
although $c_e$ increases as $\sqrt{H}$ at very low temperatures.
\end{references}
\end{document}